\documentclass[runningheads]{llncs}

\usepackage{cite}
\usepackage{fontawesome}
\usepackage{amsmath}
\usepackage{amssymb}
\usepackage[T1]{fontenc}
\usepackage{graphicx}
\usepackage{color}
\usepackage{booktabs}
\usepackage{multirow}
\usepackage{listings}
\usepackage{hyperref}
\usepackage{orcidlink}
\renewcommand{\orcidID}[1]{\,\begingroup\hypersetup{hidelinks}\orcidlink{#1}\endgroup}

\definecolor{codekeyword}{rgb}{0.13,0.29,0.53}
\definecolor{codecomment}{rgb}{0.40,0.45,0.40}
\definecolor{codestring}{rgb}{0.60,0.25,0.20}

\lstdefinestyle{scrydb}{
  language=Python,
  basicstyle=\ttfamily\scriptsize,
  keywordstyle=\color{codekeyword}\bfseries,
  commentstyle=\color{codecomment}\itshape,
  stringstyle=\color{codestring},
  showstringspaces=false,
  columns=fullflexible,
  keepspaces=true,
  breaklines=true,
  frame=tb,
  framerule=0.6pt,
  aboveskip=0.5em,
  belowskip=0.5em,
  xleftmargin=0.5em,
  emph={Index,SentenceEmbedding,scrydb},
  emphstyle=\color{codekeyword}
}
\AtBeginDocument{%
  \setlength\abovedisplayskip{2pt plus 2pt minus 1pt}%
  \setlength\belowdisplayskip{2pt plus 2pt minus 1pt}%
  \setlength\abovedisplayshortskip{2pt plus 1pt minus 1pt}%
  \setlength\belowdisplayshortskip{3pt plus 1pt minus 1pt}%
}

\begin{document}

\title{SQLite is Enough. \\ Lexical, Semantic, and Hybrid Search with \texttt{scrydb}}
\titlerunning{SQLite is Enough. Lexical, Semantic, and Hybrid Search with \texttt{scrydb}}
\author{Timo Breuer\inst{1}\orcidID{0000-0002-1765-2449}}
\authorrunning{T. Breuer}
\institute{TH Köln -- University of Applied Sciences, 50968 Köln, Germany \\
\email{timobreuer@acm.org}}
\maketitle  

\begin{abstract}
This work introduces \texttt{scrydb}, a Python library that enables lexical, semantic, and hybrid search within SQLite. For lexical search, \texttt{scrydb} leverages SQLite's full-text search extension FTS5. Semantic search builds on \texttt{sqlite-vec}, a SQLite extension for vector search. Furthermore, the library allows users to rerank and fuse retrieval results to combine both lexical and semantic approaches, providing a lightweight solution for downstream tasks in information retrieval (IR) or agentic search. We evaluate \texttt{scrydb} on various IR benchmark datasets and demonstrate its effectiveness in text retrieval based on keyword matching, semantic similarity, and rank fusion. In addition, we provide insights into query latency and the trade-off between efficiency and effectiveness. \texttt{scrydb} is available under the MIT license.

\vspace{1.75em}
\hspace{7em}
\parbox[c]{\columnwidth}
{
    \vspace{-.55em}
    \faGithub \ \href{https://github.com/breuert/scrydb/}{\nolinkurl{https://github.com/breuert/scrydb/}}
}
\vspace{-1.25em}

\end{abstract}

\section{Introduction}
\label{sec:introduction}
\texttt{scrydb} is built around the idea of packaging an entire small-to-medium-scale search pipeline and the corresponding resources, i.e., documents, lexical index, and embeddings, into an artifact that is as easy to share, archive, and rerun as a single file. It is a minimalist library for lexical, semantic, and hybrid search that stores everything a text retrieval component needs inside a single SQLite database file, with no server process, no external index, and no separate vector store to keep in sync. Lexical search is provided by SQLite's \href{https://sqlite.org/fts5.html}{FTS5} extension and its BM25 ranking function~\cite{DBLP:conf/trec/RobertsonWJHG94}. Semantic search over the entire index is enabled by \href{https://github.com/asg017/sqlite-vec}{\texttt{sqlite-vec}}, a SQLite extension for vector search at different levels of precision. For instance, binary embeddings enable efficient search, while results can also be retrieved with full-precision embeddings at the cost of increased query latency. The retrieved first-stage candidates can optionally be reranked using more costly approaches, e.g., with full-precision embeddings and cosine similarity. Alternatively, the full embeddings can be discarded to keep disk usage low. Additionally, hybrid search allows fusing the lexical and semantic rankings via Reciprocal Rank Fusion (RRF)~\cite{DBLP:conf/sigir/CormackCB09}.

The core contributions of this work are as follows:

\begin{itemize}
    \item \textbf{Lexical, semantic, and hybrid search with SQLite.} \texttt{scrydb} implements lexical scoring, semantic scoring, second-stage reranking, and rank fusion in a single library, rather than requiring application-level orchestration across multiple systems.
    \item \textbf{Single-file reproducibility with an archival-grade storage format.} By colocating raw documents, the lexical index, and embeddings in one SQLite file, \texttt{scrydb} reduces an entire IR resource to a single artifact: one file to share it all, rather than a bundle of index dumps, vector-store snapshots, and configuration files. Consequently, a \texttt{scrydb} database inherits the long-term preservation properties of any other archival dataset.
    \item \textbf{Experimental evaluation of effectiveness and efficiency.} We evaluate \texttt{scrydb} on a range of BEIR datasets~\cite{DBLP:conf/nips/Thakur0RSG21}, comparing it to MTEB baseline results~\cite{DBLP:conf/eacl/MuennighoffTMR23}. Specifically, the experiments shed light on query latency and on the effectiveness of different retrieval pipelines and levels of vector precision, providing insights into the trade-offs between retrieval quality and resource requirements.
    \item \textbf{Reusable resources for follow-up research.} \texttt{scrydb} is an open-source Python library released under the MIT license, allowing others to build on it. Additionally, we provide prebuilt \texttt{scrydb} databases for the evaluated datasets, including embeddings computed with Qwen3-Embedding-8B~\cite{DBLP:journals/corr/abs-2506-05176}, and the final retrieval runs used in our evaluation, so that future work can reproduce our results or use them as baselines for comparison.\footnote{\url{https://github.com/breuert/scrydb/} also shares pointers to the evaluation protocol and resources.}
\end{itemize}

\section{Related Work}
\label{sec:related-work}
Beyond raw effectiveness, the IR and NLP communities have increasingly scrutinized the computational cost and environmental footprint of the systems they build~\cite{DBLP:conf/sigir/ScellsZZ22,DBLP:conf/acl/StrubellGM19}. This has fueled a broader call for retrieval methods and system designs that are more efficient and less energy-intensive, rather than the pursuit of effectiveness gains at any computational cost. The ReNeuIR workshop series~\cite{DBLP:conf/sigir/BruchLN22} has become a forum for this discussion, examining efficiency in the era of neural information retrieval and encouraging more sustainable research by identifying best practices in the development of retrieval models.

Alongside efficiency, reproducibility has become a central concern for the IR community when it comes to making research practices more sustainable~\cite{Maistro2026}. \texttt{scrydb} contributes to this goal by facilitating single-file reproducibility: it relies on SQLite, which the Library of Congress lists as a preferred dataset format in its Recommended Formats Statement (\href{https://www.loc.gov/preservation/resources/rfs/data.html}{RFS}).

Neural bi-encoder models such as Dense Passage Retrieval (DPR)~\cite{DBLP:conf/emnlp/KarpukhinOMLWEC20} and Sentence-BERT~\cite{DBLP:conf/emnlp/ReimersG19} map queries and documents into a shared embedding space in which semantic relevance is approximated by vector similarity. At scale, exhaustively comparing a query embedding against every document embedding is costly, motivating a large body of work on approximate nearest neighbor (ANN) search, like inverted-file and product-quantization indexes as implemented in Faiss~\cite{DBLP:journals/tbd/JohnsonDJ21,DBLP:journals/pami/JegouDS11}, and graph-based indexes such as Hierarchical Navigable Small World (HNSW)~\cite{DBLP:journals/pami/MalkovY20}, both trading exactness for sublinear query time.

Recently, the practice of binarizing (and, less aggressively, scalar-quantizing to int8) the output of sentence-embedding models has been popularized as a way to shrink the memory footprint of embedding indexes by an order of magnitude~\cite{DBLP:books/sp/Bruch24}. The use of random hyperplane projections to derive compact binary codes from real-valued vectors traces back to locality-sensitive hashing and, in particular, SimHash~\cite{DBLP:conf/stoc/Charikar02}, which underlies sign-based binarization. Product quantization~\cite{DBLP:journals/pami/JegouDS11} offers a complementary compression strategy, decomposing a vector into subspaces that are each quantized using a small codebook rather than reduced to a single bit.

Purpose-built vector databases such as \href{https://milvus.io/}{Milvus}, \href{https://www.pinecone.io/}{Pinecone}, \href{https://weaviate.io/}{Weaviate}, and \href{https://qdrant.tech/}{Qdrant} provide managed ANN indexing and hybrid retrieval features, but typically run as standalone services with their own storage engines and operational footprints. Closer to \texttt{scrydb}'s design point are systems that add vector search to an existing relational or embedded database rather than introducing a new one: \href{https://github.com/pgvector/pgvector/}{\texttt{pgvector}} contributes vector types and ANN indexes to PostgreSQL, while DuckDB~\cite{DBLP:conf/sigmod/RaasveldtM19} follows a similar embedded, single-file philosophy for OLAP workloads and likewise offers a vector-search extension. \texttt{scrydb} shares this embedded philosophy but targets text retrieval specifically, combining lexical scoring, quantized vector search, reranking, and rank fusion into a single library and a single database file.

\section{Efficient Semantic Search Within SQLite}
\label{sec:methodology}
Semantic search becomes practically feasible within SQLite through binarization of the embeddings and comparison via the Hamming distance, with int8 scalar quantization as an intermediate precision. We refer the reader to earlier work for details on how the Hamming distance over binary embeddings approximates cosine similarity over full, 32-bit floating-point embeddings~\cite{DBLP:conf/stoc/Charikar02,DBLP:journals/jacm/GoemansW95,DBLP:books/sp/Bruch24}.

\paragraph{Full Embeddings and Cosine Similarity.}

For semantic search, each document $D$ and query $Q$ is mapped to a dense embedding vector $\mathbf{e} \in \mathbb{R}^d$, denoted $\mathbf{e}_D$ and $\mathbf{e}_Q$, respectively. The relevance of a document to a query is conventionally measured by the cosine similarity between their embeddings,
\begin{equation}
\label{eq:cosine}
\mathrm{sim}_{\cos}(\mathbf{e}_Q, \mathbf{e}_D) = \frac{\mathbf{e}_Q \cdot \mathbf{e}_D}{\lVert \mathbf{e}_Q \rVert \, \lVert \mathbf{e}_D \rVert} = \cos(\theta),
\end{equation}
where $\theta$ is the angle between $\mathbf{e}_Q$ and $\mathbf{e}_D$. Although this measure is accurate, computing it exhaustively over a full-precision, high-dimensional index is memory- and compute-intensive, motivating lower-precision representations.

\paragraph{Embedding Binarization and Hamming Distance.}

To keep both storage and comparison cost low, each $d$-dimensional embedding $\mathbf{e} = (e_1, \dots, e_d)$ is binarized into a bit vector $\mathbf{b} = (b_1, \dots, b_d) \in \{0,1\}^d$ using a Heaviside step quantizer applied component-wise,
\begin{equation}
\label{eq:binarize}
b_i = H(e_i) = \begin{cases} 1, & e_i > 0 \\ 0, & e_i \le 0 \end{cases}, \qquad i = 1, \dots, d.
\end{equation}
The resulting bit vector is packed into $d/8$ bytes and stored alongside the document, reducing the storage footprint by a factor of 32 relative to 32-bit floating-point embeddings. Semantic search over the full index is performed by comparing the binary query code $\mathbf{b}_Q$ against every binary document code $\mathbf{b}_D$ using the Hamming distance, evaluated by a bitwise XOR followed by a population count,
\begin{equation}
\label{eq:hamming}
d_H(\mathbf{b}_Q, \mathbf{b}_D) = \sum_{i=1}^{d} \left( b_{Q,i} \oplus b_{D,i} \right) = \mathrm{popcount}(\mathbf{b}_Q \oplus \mathbf{b}_D).
\end{equation}

\paragraph{Scalar Quantization to int8.}

As an intermediate point between the 1-bit code and the full 32-bit embedding, scalar quantization to int8 maps each component of an L2-normalized embedding from $[-1, 1]$ onto the int8 range by the affine rule $q_i = \mathrm{trunc}\big(255\,(e_i + 1)/2 - 128\big)$. The result is $4\times$ smaller than float32 and retains per-dimension magnitude rather than only the sign, so int8 embeddings are ranked by cosine similarity, trading some of the binary code's speed for a closer approximation of Equation~\eqref{eq:cosine}.

\section{Software Library and Implementation Details}
\label{sec:software}
\texttt{scrydb} is published on the Python Package Index and can be installed with a single command, \texttt{pip install scrydb}. Neither SQLite extension it builds on has to be compiled locally. FTS5 is part of virtually every SQLite build shipped with CPython, and \texttt{sqlite-vec} ships prebuilt loadable binaries as a pip package. Installation is a plain, wheel-only download whose only further runtime dependencies are \texttt{numpy} and \texttt{tqdm}, with the heavier machine-learning stack kept behind optional extras. The package is released under the MIT license and ships a command-line interface and a \texttt{Dockerfile}, so that it can also be driven without writing any code.

\subsection{SQLite Extensions}
\label{sec:extensions}

Lexical retrieval is delegated to \href{https://sqlite.org/fts5.html}{FTS5}, the full-text search module that is part of SQLite itself. It exposes an inverted index as a virtual table, queried through the \texttt{MATCH} operator and maintained transactionally alongside ordinary tables in the same database file. \texttt{scrydb} also surfaces FTS5's \texttt{snippet()} and \texttt{highlight()} auxiliary functions, which return short keyword-in-context fragments or fully marked-up passages without any post-processing on the application side.

\href{https://github.com/asg017/sqlite-vec}{\texttt{sqlite-vec}} stores float32, int8, or binary vectors in virtual tables queried via a dedicated \texttt{vec0} module. \texttt{scrydb} maintains one \texttt{vec0} table per collection and precision, so that binary, int8, and full-precision embeddings are stored side by side in the same database, with both the binarization and the int8 scalar quantization carried out inside SQLite. The full index can thus be scanned exhaustively at any of the three precisions, trading retrieval cost against ranking quality, and any one of them can rerank the candidates returned by another.

\subsection{Library Interface}
\label{sec:api}

The public API is centered on a single \texttt{Index} object that owns the underlying SQLite connection and doubles as a context manager, mirroring the ergonomics of \texttt{sqlite3.connect}. Figure~\ref{fig:usage} shows a complete retrieval experiment, from indexing a corpus to writing a TREC run, in a handful of lines.

\begin{figure}[!t]
\begin{lstlisting}
from scrydb import Index, SentenceEmbedding

with Index.open("scifact.db") as index:
    index.add_model(SentenceEmbedding("Qwen/Qwen3-Embedding-8B"))
    # Embeddings are computed on the fly, or taken verbatim from
    # `embedding_field` if the input rows already carry them.
    index.index_documents("corpus.jsonl", id_field="docid")
    index.index_queries("queries.jsonl", id_field="qid")

    # Interactive search: FTS5 and sqlite-vec, fused with RRF.
    hits = index.search("vitamin B12", mode="hybrid", rerank="float")
    text = index.documents[hits[0].id]["text"]   # mapping-style lookups
    emb = index.document_embeddings_binary[hits[0].id]

    # Batch retrieval over all stored queries, exported as a TREC run.
    run = index.batch_search(mode="semantic", precision="binary")
    run.write_trec("run.trec", tag="hamming")
\end{lstlisting}
\caption{Indexing a corpus and running interactive as well as batch retrieval with \texttt{scrydb}.}
\label{fig:usage}
\end{figure}

Documents and queries are indexed by providing a JSONL path or any iterable of dictionary-like rows, e.g., a Hugging Face dataset or a \texttt{pandas} data frame. \texttt{scrydb} embeds query and document texts out of the box with Sentence Transformers, the library introduced alongside Sentence-BERT~\cite{DBLP:conf/emnlp/ReimersG19}, configured through the \texttt{SentenceEmbedding} wrapper. Precomputed embeddings can also be stored directly by naming the field that carries them, decoupling \texttt{scrydb} from any particular embedding model.

\texttt{search()} takes a single query string, while \texttt{batch\_search()} evaluates all queries stored in the index. Both share the same three parameters: \texttt{mode} selects lexical, semantic, or hybrid retrieval; \texttt{precision} selects the representation of the semantic stage, namely binary, int8, or float; and \texttt{rerank} optionally adds a second stage over the top candidates of the first, at any of the three precisions.

An index also behaves like a set of read-only Python mappings: \texttt{documents} and \texttt{queries} resolve an identifier to the stored record; \texttt{document\_embeddings} and \texttt{query\_embeddings}, together with their \texttt{\_binary} and \texttt{\_int8} variants, resolve identifiers to \texttt{numpy} arrays, making a \texttt{scrydb} database a document and embedding store in its own right. Results support both attribute- and dictionary-style access, and \texttt{batch\_search()} returns a \texttt{Run} that exports to the standard six-column TREC format or to a \texttt{pandas} data frame.

\section{Experimental Evaluations}
\label{sec:experiments}
\texttt{scrydb} is evaluated on eight publicly available retrieval datasets drawn from BEIR~\cite{DBLP:conf/nips/Thakur0RSG21}, including \textbf{ArguAna}~\cite{DBLP:conf/acl/WachsmuthSS18} (8.67K documents, argument retrieval), \textbf{FiQA}~\cite{DBLP:conf/www/MaiaHFDMZB18} (57K documents, financial question-answering), \textbf{NFCorpus}~\cite{DBLP:conf/ecir/BotevaGSR16} (3.6K documents, biomedical/nutrition retrieval), \textbf{Quora}~\cite{DBLP:conf/nips/Thakur0RSG21,DBLP:conf/sigir/KamallooTLMYL24} (523K documents, duplicate question-pair retrieval), \textbf{SciDocs}~\cite{DBLP:conf/acl/CohanFBDW20} (25K documents, scientific document retrieval), \textbf{SciFact}~\cite{DBLP:conf/emnlp/WaddenLLWZCH20} (5K documents, fact-checking), \textbf{Touch\'e}~\cite{DBLP:conf/clef/BondarenkoFBGAP20a} (382K documents, argument retrieval), and \textbf{TREC-COVID}~\cite{DBLP:conf/sigir/VoorheesR21} (171K documents, ad-hoc biomedical retrieval).


As a baseline, we use the retrieval effectiveness figures reported by the Massive Text Embedding Benchmark (\textbf{MTEB})~\cite{DBLP:conf/eacl/MuennighoffTMR23}, which hosts BEIR's retrieval tasks as one of its constituent benchmark families and publishes per-model, per-dataset leaderboard results. Specifically, we take the officially reported MTEB results for Qwen3-Embedding-8B~\cite{DBLP:journals/corr/abs-2506-05176} computed with the model's full, 32-bit floating-point embeddings. This baseline represents the retrieval quality attainable by full-precision cosine search. 

For our own runs, we independently recompute the query and document embeddings for all eight BEIR datasets above using the same model, Qwen3-Embedding-8B. The following retrieval methods are evaluated.

The lexical configurations start from \textbf{BM25}~\cite{DBLP:conf/trec/RobertsonWJHG94}, returning 1000 documents per query directly from the FTS5 index. Optionally, a reranking pass over this entire candidate list is added, performed with either the Hamming distance (\textbf{BM25\,+\,Hamming}) or cosine similarity over the int8-quantized and full-precision embeddings (\textbf{BM25\,+\,cos\textsubscript{int8}} or \textbf{BM25\,+\,cos\textsubscript{float}}, respectively).

The semantic configurations instead retrieve by exhaustively scanning the entire corpus's embeddings at a given precision, returning 1000 documents per query (\textbf{Hamming}, \textbf{cos\textsubscript{int8}}, and \textbf{cos\textsubscript{float}}). Each of the two coarser first-stage rankings can optionally be refined by reranking candidates at a higher precision (\textbf{Hamming\,+\,cos\textsubscript{int8}}, \textbf{Hamming\,+\,cos\textsubscript{float}}, \textbf{cos\textsubscript{int8}\,+\,cos\textsubscript{float}}).

Finally, the hybrid configurations fuse a lexical and a semantic ranking with RRF~\cite{DBLP:conf/sigir/CormackCB09}: \textbf{RRF(Hamming)}, \textbf{RRF(cos\textsubscript{int8})}, and \textbf{RRF(cos\textsubscript{float})} each fuse the BM25 ranking with the corresponding semantic ranking, without any reranking.

All retrieval experiments are run on a single consumer machine: an Apple MacBook Air (Mac14,15) with an Apple M2 system-on-chip (8 cores: 4 performance + 4 efficiency; ARM64) and 24\,GB of unified memory, running macOS 26.5.2. No GPU is used for retrieval; the document and query embeddings are computed once, ahead of time, through a remote embedding API.

\subsection{Retrieval Effectiveness}
\label{sec:retrieval-effectiveness}

\begin{figure}[!t]
    \centering
    \includegraphics[width=\linewidth]{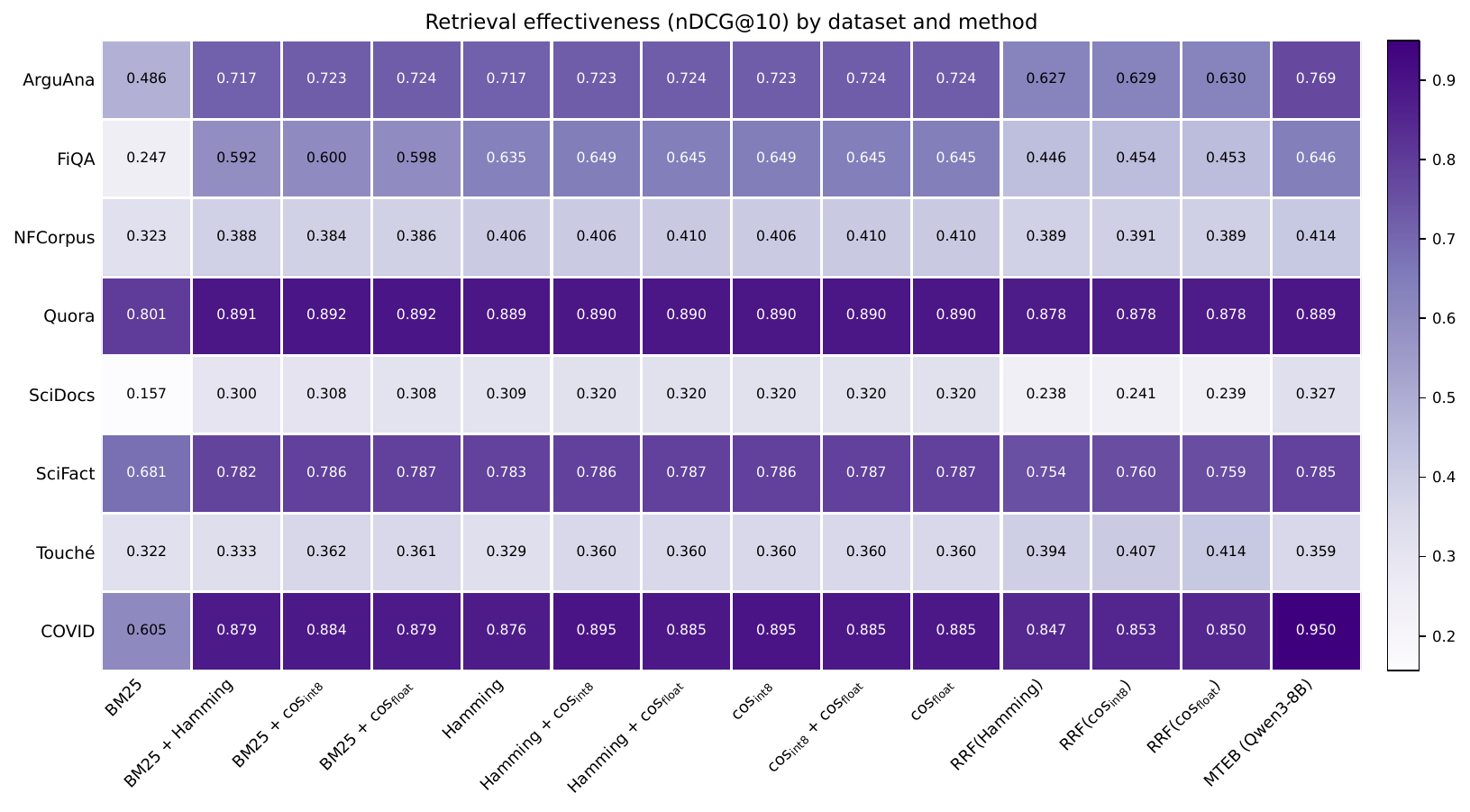}
    \caption{Retrieval effectiveness (nDCG@10) evaluated by dataset for all methods.}
    \label{fig:effectiveness_by_dataset}
\end{figure}

Table~\ref{tab:retrieval-effectiveness} reports four standard IR effectiveness measures, including Average Precision (AP), Reciprocal Rank (RR), Precision@10 (P@10), and nDCG@10, for each of the thirteen \texttt{scrydb} retrieval configurations against the full-precision MTEB baseline for Qwen3-Embedding-8B, across all eight BEIR datasets; Figure~\ref{fig:effectiveness_by_dataset} renders the same nDCG@10 figures as a dataset-by-method heatmap. We focus the discussion on nDCG@10, the primary measure used by both BEIR and MTEB, and use it as a shorthand for the general effectiveness trends, which are consistent across AP, RR, and P@10 unless noted otherwise.

\begin{table}[!t]
\centering
\resizebox{\textwidth}{!}{%
\begin{tabular}{l|l|r|r|r|r|r|r|r|r|r|r|r|r|r|r}
\toprule
Dataset & Measure & \rotatebox{90}{\shortstack[l]{BM25}} & \rotatebox{90}{\shortstack[l]{BM25 + Hamming}} & \rotatebox{90}{\shortstack[l]{BM25 + cos\textsubscript{int8}}} & \rotatebox{90}{\shortstack[l]{BM25 + cos\textsubscript{float}}} & \rotatebox{90}{\shortstack[l]{Hamming}} & \rotatebox{90}{\shortstack[l]{Hamming + cos\textsubscript{int8}}} & \rotatebox{90}{\shortstack[l]{Hamming + cos\textsubscript{float}}} & \rotatebox{90}{\shortstack[l]{cos\textsubscript{int8}}} & \rotatebox{90}{\shortstack[l]{cos\textsubscript{int8} + cos\textsubscript{float}}} & \rotatebox{90}{\shortstack[l]{cos\textsubscript{float}}} & \rotatebox{90}{\shortstack[l]{RRF(Hamming)}} & \rotatebox{90}{\shortstack[l]{RRF(cos\textsubscript{int8})}} & \rotatebox{90}{\shortstack[l]{RRF(cos\textsubscript{float})}} & \rotatebox{90}{\shortstack[l]{MTEB (Qwen3-8B)}} \\
\midrule
\multirow{4}{*}{ArguAna} & AP & 0.405 & 0.643 & 0.648 & \underline{0.649} & 0.643 & 0.648 & 0.649 & 0.648 & 0.649 & 0.649 & 0.548 & 0.549 & 0.551 & \textbf{0.705} \\
 & RR & 0.405 & 0.643 & 0.648 & \underline{0.649} & 0.643 & 0.648 & 0.649 & 0.648 & 0.649 & 0.649 & 0.548 & 0.549 & 0.551 & \textbf{0.705} \\
 & P@10 & 0.078 & 0.095 & 0.096 & 0.096 & 0.095 & 0.096 & \underline{0.096} & 0.096 & \underline{0.096} & \underline{0.096} & 0.089 & 0.090 & 0.090 & \textbf{0.097} \\
 & nDCG@10 & 0.486 & 0.717 & 0.723 & \underline{0.724} & 0.717 & 0.723 & 0.724 & 0.723 & 0.724 & 0.724 & 0.627 & 0.629 & 0.630 & \textbf{0.769} \\
\midrule
\multirow{4}{*}{FiQA} & AP & 0.205 & 0.525 & 0.530 & 0.529 & 0.567 & 0.578 & 0.575 & \underline{0.578} & 0.575 & 0.575 & 0.383 & 0.392 & 0.390 & \textbf{0.578} \\
 & RR & 0.314 & 0.684 & 0.693 & 0.689 & 0.711 & \textbf{0.726} & 0.721 & \underline{0.726} & 0.721 & 0.721 & 0.533 & 0.543 & 0.542 & 0.724 \\
 & P@10 & 0.068 & 0.155 & 0.158 & 0.158 & 0.173 & \textbf{0.177} & \underline{0.176} & \textbf{0.177} & \underline{0.176} & \underline{0.176} & 0.126 & 0.127 & 0.127 & 0.175 \\
 & nDCG@10 & 0.247 & 0.592 & 0.600 & 0.598 & 0.635 & \textbf{0.649} & 0.645 & \textbf{0.649} & 0.645 & 0.645 & 0.446 & 0.454 & 0.453 & \underline{0.646} \\
\midrule
\multirow{4}{*}{NFCorpus} & AP & 0.151 & 0.182 & 0.181 & 0.182 & 0.217 & 0.220 & 0.222 & 0.220 & 0.222 & \underline{0.222} & 0.207 & 0.212 & 0.212 & \textbf{0.225} \\
 & RR & 0.525 & 0.615 & 0.604 & 0.606 & 0.620 & 0.616 & 0.620 & 0.616 & 0.620 & \underline{0.620} & 0.604 & 0.606 & 0.603 & \textbf{0.627} \\
 & P@10 & 0.231 & 0.284 & 0.280 & 0.282 & 0.301 & 0.300 & \underline{0.304} & 0.300 & \underline{0.304} & \underline{0.304} & 0.288 & 0.287 & 0.286 & \textbf{0.308} \\
 & nDCG@10 & 0.323 & 0.388 & 0.384 & 0.386 & 0.406 & 0.406 & \underline{0.410} & 0.406 & \underline{0.410} & \underline{0.410} & 0.389 & 0.391 & 0.389 & \textbf{0.414} \\
\midrule
\multirow{4}{*}{Quora} & AP & 0.759 & 0.861 & \underline{0.862} & \textbf{0.862} & 0.859 & 0.861 & 0.861 & 0.861 & 0.861 & 0.861 & 0.844 & 0.845 & 0.845 & 0.859 \\
 & RR & 0.797 & 0.882 & \underline{0.883} & \textbf{0.883} & 0.880 & 0.881 & 0.881 & 0.881 & 0.881 & 0.881 & 0.873 & 0.874 & 0.874 & 0.879 \\
 & P@10 & 0.121 & 0.135 & \underline{0.136} & \textbf{0.136} & 0.135 & 0.136 & 0.136 & 0.136 & 0.136 & 0.136 & 0.133 & 0.133 & 0.133 & 0.136 \\
 & nDCG@10 & 0.801 & 0.891 & \underline{0.892} & \textbf{0.892} & 0.889 & 0.890 & 0.890 & 0.890 & 0.890 & 0.890 & 0.878 & 0.878 & 0.878 & 0.889 \\
\midrule
\multirow{4}{*}{SciDocs} & AP & 0.109 & 0.216 & 0.221 & 0.221 & 0.236 & 0.243 & 0.244 & 0.243 & 0.244 & \underline{0.244} & 0.182 & 0.185 & 0.184 & \textbf{0.252} \\
 & RR & 0.285 & 0.483 & 0.492 & 0.492 & 0.495 & 0.500 & \underline{0.503} & 0.500 & 0.503 & 0.503 & 0.414 & 0.415 & 0.413 & \textbf{0.510} \\
 & P@10 & 0.081 & 0.156 & 0.161 & 0.161 & 0.162 & \underline{0.170} & 0.169 & \underline{0.170} & 0.169 & 0.169 & 0.123 & 0.124 & 0.123 & \textbf{0.173} \\
 & nDCG@10 & 0.157 & 0.300 & 0.308 & 0.308 & 0.309 & \underline{0.320} & 0.320 & \underline{0.320} & 0.320 & 0.320 & 0.238 & 0.241 & 0.239 & \textbf{0.327} \\
\midrule
\multirow{4}{*}{SciFact} & AP & 0.642 & 0.740 & \underline{0.745} & \textbf{0.745} & 0.741 & 0.743 & 0.744 & 0.743 & 0.744 & 0.744 & 0.710 & 0.718 & 0.717 & 0.741 \\
 & RR & 0.653 & 0.748 & 0.752 & \textbf{0.752} & 0.749 & 0.751 & 0.752 & 0.751 & \underline{0.752} & \underline{0.752} & 0.720 & 0.730 & 0.729 & 0.749 \\
 & P@10 & 0.089 & 0.103 & 0.104 & 0.104 & 0.104 & \underline{0.104} & \textbf{0.105} & \underline{0.104} & \textbf{0.105} & \textbf{0.105} & 0.099 & 0.099 & 0.099 & 0.104 \\
 & nDCG@10 & 0.681 & 0.782 & 0.786 & \underline{0.787} & 0.783 & 0.786 & \textbf{0.787} & 0.786 & \textbf{0.787} & \textbf{0.787} & 0.754 & 0.760 & 0.759 & 0.785 \\
\midrule
\multirow{4}{*}{Touché} & AP & 0.229 & 0.238 & 0.247 & 0.248 & 0.233 & 0.243 & 0.244 & 0.243 & 0.244 & 0.244 & 0.278 & \textbf{0.291} & \underline{0.290} & 0.254 \\
 & RR & 0.600 & 0.588 & 0.652 & 0.621 & 0.587 & 0.649 & 0.618 & 0.649 & 0.618 & 0.618 & 0.680 & \underline{0.682} & \textbf{0.700} & 0.576 \\
 & P@10 & 0.294 & 0.294 & 0.316 & 0.316 & 0.288 & 0.314 & 0.316 & 0.314 & 0.316 & 0.316 & 0.343 & \underline{0.347} & \textbf{0.355} & 0.318 \\
 & nDCG@10 & 0.322 & 0.333 & 0.362 & 0.361 & 0.329 & 0.360 & 0.360 & 0.360 & 0.360 & 0.360 & 0.394 & \underline{0.407} & \textbf{0.414} & 0.359 \\
\midrule
\multirow{4}{*}{COVID} & AP & 0.195 & 0.303 & 0.308 & 0.308 & 0.393 & 0.407 & 0.406 & \underline{0.413} & 0.412 & 0.412 & 0.358 & 0.367 & 0.366 & \textbf{0.484} \\
 & RR & 0.885 & 0.975 & \textbf{1.000} & \textbf{1.000} & 0.985 & \textbf{1.000} & \underline{0.990} & \textbf{1.000} & \underline{0.990} & \underline{0.990} & 0.977 & 0.970 & 0.967 & \textbf{1.000} \\
 & P@10 & 0.636 & 0.918 & 0.916 & 0.910 & 0.912 & \underline{0.926} & 0.916 & \underline{0.926} & 0.916 & 0.916 & 0.882 & 0.890 & 0.890 & \textbf{0.974} \\
 & nDCG@10 & 0.605 & 0.879 & 0.884 & 0.879 & 0.876 & \underline{0.895} & 0.885 & \underline{0.895} & 0.885 & 0.885 & 0.847 & 0.853 & 0.850 & \textbf{0.950} \\
\bottomrule
\end{tabular}%
}
\caption{Retrieval effectiveness (AP, RR, P@10, nDCG@10) of each retrieval method against the full-precision embedding baseline reported by MTEB for Qwen3-Embedding-8B. Best score per row in \textbf{bold}, second-best \underline{underlined}; tied methods share the mark, with ties determined by the unrounded scores.}
\label{tab:retrieval-effectiveness}
\end{table}

Reranking improves over raw BM25 in nDCG@10 on every dataset, substantially so on all but Touch\'e, where the gain is marginal and RR in fact degrades slightly. Notably, reranking by the inexpensive Hamming distance already captures nearly all of this gain. The binary codes therefore retain enough of the relevance signal to serve as a cheap drop-in for cosine-based reranking on most, though not all, corpora. Semantic retrieval over the full binarized index outperforms BM25 in nDCG@10 on all eight datasets, in several cases by a wide margin, confirming that the semantic signal captured by the binarized Qwen3-Embedding-8B embeddings is more informative than exact lexical overlap.

A more consequential pattern emerges from the reranked semantic configurations: refining a coarse ranking to higher precision is equivalent, at the top of the ranking, to scanning the whole corpus at that precision. Hamming\,+\,cos\textsubscript{int8} and exhaustive cos\textsubscript{int8} yield identical P@10 and nDCG@10 to three decimals on all eight datasets, as do Hamming\,+\,cos\textsubscript{float} and cos\textsubscript{float}, and likewise cos\textsubscript{int8}\,+\,cos\textsubscript{float} and cos\textsubscript{float}; within each pair the columns diverge only in AP, where documents ranked below the 1000-document reranking depth still contribute. The binary first stage thus already places the documents that a full-precision exhaustive scan would rank in the top ten inside its top-1000 candidates, so the coarse Hamming pass costs nothing in early-precision terms. Since rescoring 1000 candidates is far cheaper than scanning an entire corpus at int8 or float precision, Hamming\,+\,cos\textsubscript{int8} dominates exhaustive cos\textsubscript{int8} at the top of the ranking, matching its top-10 effectiveness at roughly one-fifth of the mean latency (164.5\,ms vs.\ 822.5\,ms) while trailing it marginally in AP (0.407 vs.\ 0.413 on TREC-COVID); by the same argument Hamming\,+\,cos\textsubscript{float} dominates plain cos\textsubscript{float}, being cheaper and indistinguishable in terms of nDCG@10. This is the central effectiveness argument for the binary-first design.

On four of the eight datasets, a \texttt{scrydb} configuration meets or exceeds the MTEB baseline: Hamming\,+\,cos\textsubscript{int8} reaches 0.649 nDCG@10 on FiQA vs.\ 0.646, BM25\,+\,cos\textsubscript{float} 0.892 on Quora vs.\ 0.889, Hamming\,+\,cos\textsubscript{float} 0.787 on SciFact vs.\ 0.785, and RRF(cos\textsubscript{float}) 0.414 on Touch\'e vs.\ 0.359. 
The remaining gaps in effectiveness are not attributable to \texttt{scrydb}'s compressed representations. Our own exhaustive full-precision run, cos\textsubscript{float}, which performs precisely the computation the baseline describes, reproduces the TREC-COVID deficit (0.885 nDCG@10 against the reported 0.950), so the deficit is already present before any binarization or quantization is applied and more likely originates in the embedding pipeline, e.g., in prompt formatting, than in the index.

RRF hybrid search is the best-performing \texttt{scrydb} configuration on exactly one dataset, Touch\'e, where RRF(cos\textsubscript{float}) wins on RR, P@10, and nDCG@10 and exceeds the MTEB baseline, with RRF(cos\textsubscript{int8}) marginally ahead on AP. On the remaining seven datasets, RRF trails the better of the two rankings it fuses. RRF in \texttt{scrydb} is therefore worthwhile only where lexical and semantic retrieval are of comparable strength and make complementary rather than redundant errors, and is not a uniformly safe default.

\subsection{Query Latency}
\label{sec:query-latency}

\begin{figure}[!t]
    \centering
    \includegraphics[width=\linewidth]{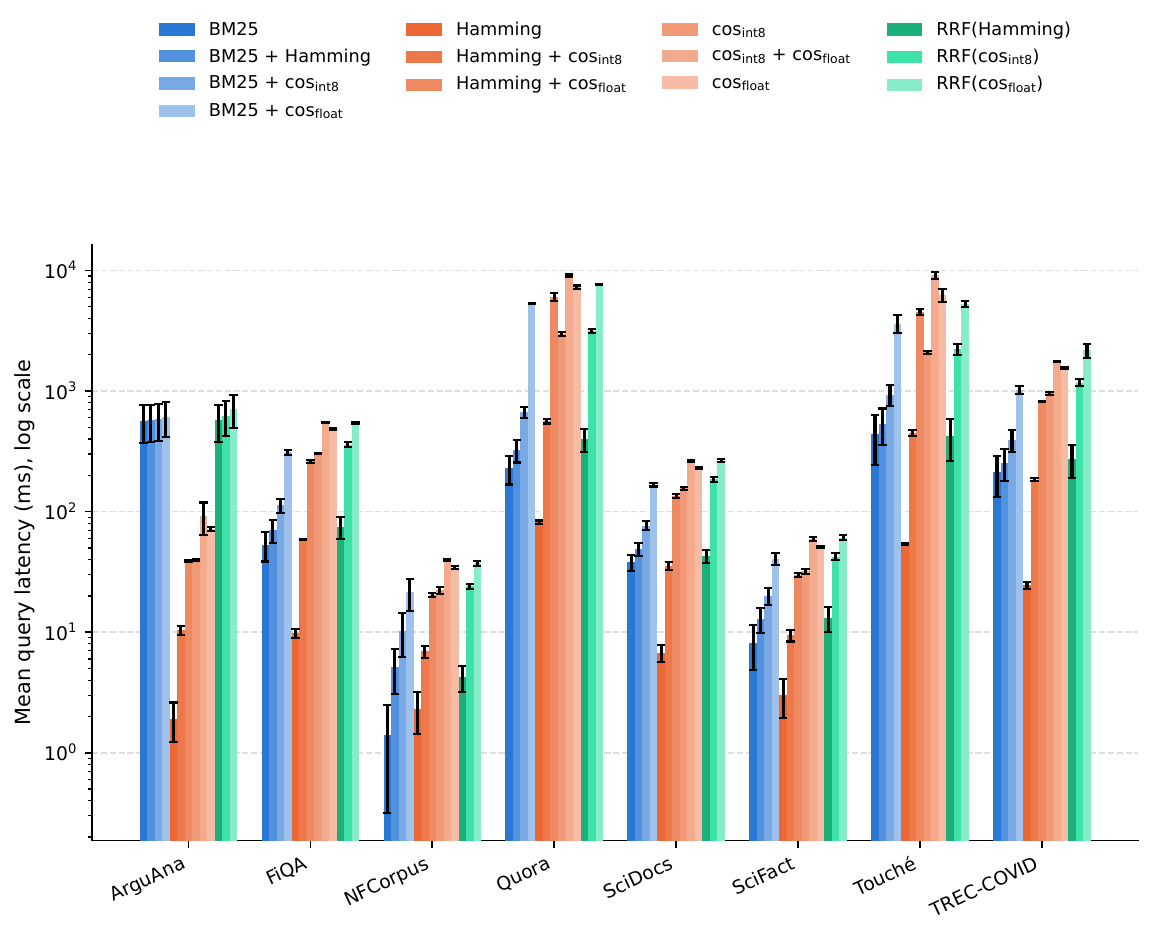}
    \caption{Query latency evaluated by dataset for all retrieval methods.}
    \label{fig:efficiency_by_dataset}
\end{figure}

Figure~\ref{fig:efficiency_by_dataset} and Table~\ref{tab:retrieval-latency} report the mean end-to-end query latency, in milliseconds, for each configuration across all eight BEIR datasets, alongside the corpus sizes and mean query lengths that relate cost to both dimensions; the figure additionally shows the standard deviation across queries as error bars and uses a logarithmic axis, since latencies span almost four orders of magnitude, from 1.4\,ms (BM25 on NFCorpus) to 9138.4\,ms (cos\textsubscript{int8}\,+\,cos\textsubscript{float} on Touch\'e).

\begin{table}[!t]
\centering
\resizebox{\textwidth}{!}{%
\begin{tabular}{l|r|r|r|r|r|r|r|r|r|r|r|r|r|r|r}
\toprule
Dataset & Size & \rotatebox{90}{Query Length} & \rotatebox{90}{\shortstack[l]{BM25}} & \rotatebox{90}{\shortstack[l]{BM25 + Hamming}} & \rotatebox{90}{\shortstack[l]{BM25 + cos\textsubscript{int8}}} & \rotatebox{90}{\shortstack[l]{BM25 + cos\textsubscript{float}}} & \rotatebox{90}{\shortstack[l]{Hamming}} & \rotatebox{90}{\shortstack[l]{Hamming + cos\textsubscript{int8}}} & \rotatebox{90}{\shortstack[l]{Hamming + cos\textsubscript{float}}} & \rotatebox{90}{\shortstack[l]{cos\textsubscript{int8}}} & \rotatebox{90}{\shortstack[l]{cos\textsubscript{int8} + cos\textsubscript{float}}} & \rotatebox{90}{\shortstack[l]{cos\textsubscript{float}}} & \rotatebox{90}{\shortstack[l]{RRF(Hamming)}} & \rotatebox{90}{\shortstack[l]{RRF(cos\textsubscript{int8})}} & \rotatebox{90}{\shortstack[l]{RRF(cos\textsubscript{float})}} \\
\midrule
ArguAna & 8.67K & 181.4 & 566.8 & 573.6 & 584.5 & 612.3 & \textbf{1.9} & \underline{10.4} & 38.7 & 40.0 & 91.3 & 71.7 & 574.7 & 626.9 & 712.7 \\
FiQA & 57K & 11.0 & \underline{53.1} & 70.7 & 112.6 & 310.7 & \textbf{9.8} & 58.5 & 261.3 & 304.9 & 547.8 & 484.1 & 74.7 & 360.3 & 543.6 \\
NFCorpus & 3.6K & 3.9 & \textbf{1.4} & 5.2 & 10.3 & 21.4 & \underline{2.3} & 6.9 & 20.3 & 22.2 & 39.6 & 34.3 & 4.2 & 24.0 & 37.2 \\
Quora & 523K & 9.9 & \underline{228.7} & 324.7 & 670.9 & 5293.5 & \textbf{81.5} & 560.6 & 6034.1 & 2971.0 & 9073.1 & 7294.8 & 397.4 & 3160.0 & 7669.6 \\
SciDocs & 25K & 10.0 & 38.0 & 48.7 & 76.5 & 167.7 & \textbf{6.7} & \underline{35.4} & 135.6 & 156.1 & 264.3 & 231.2 & 42.7 & 185.1 & 264.7 \\
SciFact & 5K & 12.6 & \underline{8.2} & 12.8 & 20.1 & 40.8 & \textbf{3.0} & 9.4 & 30.0 & 31.9 & 59.1 & 51.3 & 13.1 & 42.2 & 60.8 \\
Touché & 382K & 6.3 & 436.5 & 538.2 & 933.9 & 3635.6 & \textbf{53.7} & 449.8 & 4532.4 & 2097.8 & 9138.4 & 6259.0 & \underline{423.5} & 2225.4 & 5325.8 \\
COVID & 171K & 9.5 & 211.6 & 253.4 & 392.2 & 1020.3 & \textbf{24.4} & \underline{184.6} & 812.7 & 955.8 & 1748.2 & 1558.4 & 273.2 & 1172.8 & 2174.9 \\
\midrule
\textbf{Mean} & -- & -- & 193.0 & 228.4 & 350.1 & 1387.8 & \textbf{22.9} & \underline{164.5} & 1483.1 & 822.5 & 2620.2 & 1998.1 & 225.4 & 974.6 & 2098.7 \\
\bottomrule
\end{tabular}%
}
\caption{Mean query latency (ms) of each retrieval strategy, per BEIR dataset and averaged across datasets (\textbf{Mean} row), alongside corpus size and mean query length (words, computed over the queries timed for that row). Fastest method per row in \textbf{bold}, second-fastest \underline{underlined}.}
\label{tab:retrieval-latency}
\end{table}

Hamming search is the cheapest form of semantic retrieval and the fastest of all thirteen configurations on seven of the eight datasets. Its latency scales with corpus size rather than query length, as expected from its exhaustive, word-parallel scan over every stored binary code, growing to 81.5\,ms on the 523K-document Quora corpus. Scanning the same embeddings at higher precision is markedly more expensive: exhaustive cosine similarity over int8-quantized embeddings costs, on average across datasets, $36\times$ as much as Hamming (822.5\,ms vs.\ 22.9\,ms), and over full-precision embeddings a further $2.4\times$ beyond that (1998.1\,ms). Binarization is therefore not merely a storage optimization but the main lever that keeps exhaustive semantic search cheap.

BM25 shows the opposite dependency: because FTS5 only touches the postings lists of the terms present in a query, its latency tracks query length far more than corpus size. This is most visible on ArguAna, whose 8.67K-document corpus is the third-smallest of the eight, yet whose queries are themselves full arguments averaging 181.4 words, an order of magnitude longer than any other dataset's, so that each query touches far more postings lists than a handful of keywords would; its BM25 latency (566.8\,ms) accordingly exceeds that of the 382K-document Touch\'e corpus (436.5\,ms).

Reranking adds a cost that grows with both the underlying corpus size and the precision of the target representation, rather than staying fixed at the reranking depth of 1000 as a pure top-$k$ rescoring would suggest. Rescoring the top-1000 candidates by the compact binary codes adds roughly 0.2--1.1\,ms per thousand corpus documents on top of the first-stage ranking, by int8-quantized embeddings roughly 0.9--2.5\,ms, and by full-precision embeddings roughly 4--12\,ms (+45.5\,ms on ArguAna, but +5064.8\,ms on Quora). This ordering, which holds whether the first-stage ranking being refined is BM25's or Hamming's, tracks the on-disk width of each representation, suggesting that a reranking pass is dominated by the cost of reading the candidates' stored vectors rather than by the comparison itself.

The most expensive configurations combine an exhaustive first-stage semantic scan with a further reranking or fusion pass. Stacking a full int8 scan with a full-precision cosine rerank (cos\textsubscript{int8}\,+\,cos\textsubscript{float}) is the costliest of all, averaging 2620.2\,ms across the eight datasets and reaching 9138.4\,ms on Touch\'e, and fusing BM25 with a full-precision scan is comparably expensive. The cheaper RRF(Hamming) fusion, by contrast, broadly tracks the sum of its two standalone components plus a fusion overhead, reflecting that RRF is computed over the two raw, unreranked result lists rather than over already-reranked ones.

Despite this wide range, the cheapest configurations remain practical at scale: plain Hamming search still answers a query in 81.5\,ms at the 523K-document scale of Quora, and its int8-precision rerank averages 164.5\,ms across datasets, while the multi-second latencies are confined to configurations that stack two exhaustive high-precision scans, on the order of $100\times$ more expensive than plain Hamming search, without a correspondingly large effectiveness gain over cheaper alternatives. The word-parallel Hamming-distance scan thus keeps exhaustive binary-code search practical at this scale without resorting to approximate nearest-neighbor structures.

\section{Discussion}
\label{sec:discussion}
The experimental results put a concrete number on the central design bet behind \texttt{scrydb}: that collapsing a retrieval experiment into a single, self-contained SQLite file need not come at a prohibitive cost in either effectiveness or latency. Across the eight datasets, the best-performing \texttt{scrydb} configuration trails the full-precision MTEB baseline for Qwen3-Embedding-8B by a mean of 0.006 nDCG@10, and meets or exceeds it on four of the eight, despite operating on embeddings binarized to one thirty-second of their original size.

None of this is an argument that \texttt{scrydb} obsoletes the vector-database and ANN ecosystem. Purpose-built systems and the ANN algorithms underlying them, e.g., inverted-file and product-quantization indexes as in Faiss, and graph-based indexes such as HNSW, solve problems that a single embedded SQLite file, by construction, does not attempt to solve: concurrent, multi-tenant query serving; horizontal sharding and replication across machines; real-time ingestion and updates against a corpus that grows without bound; GPU-accelerated index construction and search; and the operational tooling needed to run all of this reliably in production. \texttt{scrydb}'s exhaustive scan over binary codes is not a substitute for the sublinear query time these systems achieve at web scale. If anything, \texttt{scrydb} is only possible because of this existing technology stack, not in spite of it.

\texttt{scrydb} is best understood as occupying a different point in the same design space rather than as a competitor to those systems: it trades away sublinear query time, horizontal scalability, and concurrent serving in exchange for being expressible as a single, dependency-light, portable file. For a production search service fielding concurrent traffic over a continuously growing corpus, a purpose-built vector database remains the right tool. This is a call to weigh that cost against the actual requirements rather than pay it unconditionally for headroom that may never be used, not a claim that lightweight, self-contained tools are categorically preferable to mature systems.

That trade-off has a concrete boundary, which the latency results let us quantify rather than merely assert. Across the eight evaluated corpora, Hamming search latency scales at a marginal rate of roughly 0.15\,ms per thousand documents, on top of a fixed per-query overhead of about 1.3\,ms. This scan does not, however, become sublinear the way an ANN index does: extrapolating the measured rate linearly, a corpus of a few million documents is still answered in under a second, but mean per-query latency reaches several seconds at around twenty million documents, and a corpus at the scale purpose-built vector databases are designed for, tens of millions to billions of documents, would be impractical to search exhaustively on commodity hardware, regardless of how cheap the word-parallel popcount makes each individual comparison.

The same $32\times$ compression that keeps this scan cheap in absolute terms also keeps its memory footprint low, and it is worth stating that footprint concretely: a 4096-dimensional Qwen3-Embedding-8B vector occupies 512 bytes once binarized, compared with 16\,KB in full 32-bit precision. Even a several-million-document collection's entire semantic index therefore amounts to a few gigabytes and fits comfortably in the memory of a consumer machine, rather than the tens of gigabytes the same collection would require at full precision. This is precisely the regime in which \texttt{scrydb} is a practical resource: for small-to-medium-scale collections, up to roughly a few million documents, it offers efficient, low-memory-footprint search that scans the entire collection exhaustively, without needing to build, tune, or maintain an approximate index at all.

\section{Conclusion}
\label{sec:conclusion}
This paper presented \texttt{scrydb}, a minimalist library that packages lexical, semantic, and hybrid search into a single, self-contained SQLite file. By colocating documents, the lexical index, and embeddings in one file, \texttt{scrydb} turns an entire retrieval resource into a single artifact that can be shared, archived, and rerun as easily as any other dataset file, making retrieval experiments more reproducible.

\texttt{scrydb} is not a replacement for the vector-database and ANN ecosystem; rather, it makes the case for reaching for that heavier, more scalable stack when a task's scale and serving requirements actually call for it, and for settling on a single, self-contained, easily archived file otherwise. We hope that this positioning makes \texttt{scrydb} databases practical artifacts for accompanying published retrieval experiments and for supporting small-to-medium-scale projects.

\bibliographystyle{splncs04}
\bibliography{references}

\end{document}